\documentclass[aps,prb,preprint,groupedaddress,showpacs]{revtex4-1}

\usepackage{graphicx}
\usepackage{epsfig}
\usepackage{dcolumn}
\usepackage{bm}
\usepackage{tabularx}
\usepackage{hyperref}
\usepackage{xcolor}
\usepackage{amsmath}
\usepackage{subfigure}
\usepackage{float}
\hypersetup{citecolor=black, filecolor= black, linkcolor= black, urlcolor= black}

\begin{document}

\title{Theoretical investigation of the pressure induced novel structure, phase transition, mechanical and electronic properties in Hf-O system.}
\title{Pressure induced novel compounds in the Hf-O system from first-principles calculations}
\author{Jin Zhang}
\email{Jin.Zhang.1@stonybrook.edu}
\affiliation{Department of Geosciences, Center for Materials by Design, and Institute for Advanced Computational Science, State University of New York, Stony Brook, NY 11794-2100, USA}
\author{Artem R. Oganov}
\email{artem.oganov@stonybrook.edu}
\affiliation{Skolkovo Institute of Science and Technology, Skolkovo Innovation Center, 5 Nobel St., Moscow, 143026, Russia.}
\affiliation{Department of Geosciences, Center for Materials by Design, and Institute for Advanced Computational Science, State University of New York, Stony Brook, NY 11794-2100, USA}
\affiliation{Science and Technology on Thermostructural Composite Materials Laboratory, International Center for Materials Discovery, School of Materials Science and Engineering, Northwestern Polytechnical University, Xi'an, Shaanxi 710072, PR China}
\affiliation{Moscow Institute of Physics and Technology, Dolgoprudny, Moscow Region 141700, Russia}

\author{Xinfeng Li}
\affiliation{State Key Laboratory for Mechanical Behavior of Materials, School of Materials Science and Engineering, Xi'an Jiaotong University, Xi¡¯an, 710049, PR China}

\author{Kan-Hao Xue}
\affiliation{School of Optical and Electronic Information, Huazhong University of Science and Technology, Wuhan 430074, China}

\author{Zhenhai Wang}
\affiliation{Peter Gr{\"u}nberg Research Center, Nanjing University of Posts and Telecommunications, Nanjing, Jiangsu 210003, China}

\author{Huafeng Dong}
\affiliation{College of Physics and Optoelectronic Engineering, Guangdong University of Technology, Guangzhou 510006, China}

\begin{abstract}

Using first-principles evolutionary simulations, we have systematically investigated phase stability in the Hf-O system at pressure up to 120 GPa. New compounds Hf$_5$O$_2$, Hf$_3$O$_2$, HfO and HfO$_3$ are discovered to be thermodynamically stable at certain pressure ranges and a new stable high-pressure phase is found for Hf$_2$O with space group  \emph{Pnnm} and anti-CaCl$_2$-type structure. Both \emph{P}$\bar{6}$2\emph{m}-HfO and \emph{P}$\bar{4}$\emph{m}2-Hf$_2$O$_3$ show semimetallic character. \emph{Pnnm}-HfO$_3$ shows interesting structure, simultaneously containing oxide O$^{2-}$ and peroxide [O-O]$^{2-}$ anions. Remarkably, it is \emph{P}$\bar{6}$2\emph{m}-HfO rather than O\uppercase\expandafter{\romannumeral2}-HfO$_2$ that exhibits the highest  mechanical characteristics among Hf-O compounds. \emph{Pnnm}-Hf$_2$O, \emph{Imm}2-Hf$_5$O$_2$, \emph{P}$\bar{3}$1\emph{m}-Hf$_{2}$O and \emph{P}$\bar{4}$\emph{m}2-Hf$_2$O$_3$ phases also show superior mechanical properties, these phases can be quenched to ambient pressure and their properties can be exploited.

\end{abstract}

\maketitle
\section{Introduction}

Hafnium oxide HfO$_{2}$ has a wide range of technological applications. In electronics industry, hafnium oxide-based material is currently used as an excellent high-\emph{k} gate dielectric\cite{choi2011development} and oxygen-deficient hafnium oxide also received additional interest for resistive-switching memories\cite{lin2011electrode}. As for other applications, even though the hardness of hafnia (HfO$_2$) is not that high for it to be considered as a superhard material\cite{al2010phase}, it still attracts attention as a potential candidate for hard oxide-based materials\cite{chung2003superhard}. Unlike carbides or nitrides, oxides are more stable in the oxygen atmosphere at high temperature, which is valuable for many applications, Many oxide ceramics, especially those involving transition metals, are promising for application as hard coatings, since metal d electrons and strong bonds define their remarkable mechanical properties (high hardness, good chemical resistance, high tensile strength, and good fracture toughness)\cite{lowther2003superhard}. As compared to most transition metal oxide ceramics, hafnium oxide ceramics exhibit enhanced mechanical properties (higher fracture toughness) and structural stability (low thermal conductivity). Here, we want to explore all possible stable compounds of Hf-O system at pressure up to 120 GPa.

Under ambient temperature, experiments\cite{xia1990crystal,ahuja1993crystal} indicated that pure Hf is stable in the $\alpha$-phase (hexagonal close-packed structure, space group: \emph{P}6$_{3}$/\emph{mmc}) and transforms to $\omega$-phase (hexagonal structure, space group: \emph{P}6/\emph{mmm}) at 46-58 GPa and then to $\beta$-Hf (body centered cubic, space group: \emph{Im}$\bar{3}$\emph{m}) at 71.1 GPa-78.4 GPa. Our GGA calculated results indicate the transition pressures: $\alpha$-phase $\rightarrow$ $\omega$-phase at 49 GPa and $\omega$-phase $\rightarrow$ $\beta$-Hf at 70 GPa, which are in accord with above experimental results. It has been suggested that the solubility of oxygen in the octahedral interstitial sites of $\alpha$-Hf (hcp-Hf) can be as high as 20 at.\%\cite{tsuji1997thermochemistry}, while solubility of oxygen in $\beta$-Hf (bcc-Hf) is only 3 at.\%\cite{massalski1990binary}. Several experimental\cite{hirabayashi1973superstructure,hirabayashi1972interstitial} and theoretical studies\cite{ruban2010oxygen,paul2011first} have investigated the interstitial oxygen in hcp-Hf. Now it is well established that three stoichiometric compositions Hf$_{6}$O, Hf$_3$O and Hf$_{2}$O can be formed with increasing occupation of the octahedral-interstitial positions in hcp-Hf by oxygen atoms. Hf$_{2}$O$_{3}$ was theoretically predicted to form upon increasing the concentration of oxygen vacancies in monoclinic HfO$_{2}$\cite{xue2013prediction}.

The phase sequence of HfO$_2$ at ambient temperature with increasing pressure is: baddeleyite (monoclinic, space group: \emph{P}2$_{1}$/\emph{c}) $\rightarrow$ orthorhombic \uppercase\expandafter{\romannumeral1} (orthorhombic, space group: \emph{Pbca}, O\uppercase\expandafter{\romannumeral1}) $\rightarrow$ orthorhombic \uppercase\expandafter{\romannumeral2} (orthorhombic, space group: \emph{Pnma}, O\uppercase\expandafter{\romannumeral2})\cite{desgreniers1999high,tang1998high,kang2003first}. Orthorhombic O\uppercase\expandafter{\romannumeral2}-HfO$_2$ with experimentally reported hardness between 6-13 GPa\cite{adams1991x} has been speculated to be much harder than the low-pressure phases (baddeleyite and O\uppercase\expandafter{\romannumeral1}-HfO$_2$) because of its comparatively high bulk modulus\cite{lowther2003superhard,ohtaka2001phase}.

In this study, we systematically investigate the structure and stability of Hf-O compounds up to a pressure of 120 GPa by the first-principles evolutionary algorithm USPEX. Several new stoichiometries in the Hf-O system have been predicted under high pressure. Furthermore, we verify the dynamical and mechanical stability of these new high-pressure phases at 0 GPa by calculating their phonons and elastic constants. To better understand the correlations between hardness and O content, we estimate the hardness of these phases at 0 GPa using Chen's hardness model\cite{chen2011modeling}. Quenchable high-pressure phases often possess superior mechanically properties, and we indeed find novel hafnium oxides with unusual mechanical properties.

\section{Computational methodology}

Searching the stable high-pressure structures in Hf-O system was done using first-principles evolutionary algorithm (EA) as implemented in the USPEX code\cite{oganov2006crystal,lyakhov2013new,oganov2011evolutionary} combined with ab initio structure relaxations using density functional theory (DFT) with the PBE-GGA functional\cite{perdew1996generalized}, as implemented in the VASP package\cite{kresse1996efficient}. In our work, variable-composition structure searches\cite{oganov2011evolutionary} for the Hf-O system with up to 20 atoms in the unit cell were performed at 0 GPa, 10 GPa, 20 GPa, 30 GPa, 40 GPa, 50 GPa, 60 GPa, 70 GPa, 80 GPa, 90 GPa, 100 GPa, 110 GPa and 120 GPa. The initial generation of structures was produced randomly using space group symmetry, each subsequent generation was obtained by variation operators including heredity (40\%), lattice mutation (20\%), random (20\%) and transmutation (20\%). The electron-ion interaction was described by the projector-augmented wave (PAW) pseudopotentials\cite{blochl1994projector}, with 5\emph{p}$^6$6\emph{s}$^2$5\emph{d}$^4$ and 2\emph{s}$^2$2\emph{p}$^4$ shells treated as valence for Hf and O, respectively. The generalized gradient approximation (GGA) in the Perdew-Burke-Ernzerhof form\cite{perdew1996generalized} was utilized for describing exchange-correlation effects. The plane-wave energy cutoff was chosen as 600 eV and $\Gamma$-centered uniform \emph{k}-meshes with resolution 2$\pi\times0.06$ {\AA}$^{-1}$ were used to sample the Brillouin zone, resulting in excellent convergence. Phonon dispersions were calculated using the finite-displacement method with the Phonopy code\cite{togo2008first}.

\section{Results and discussions}
\subsection{Crystal structure prediction for the Hf-O system}

\begin{center}
\begin{figure}[h]
   \includegraphics[angle=0,width=1.0\linewidth]{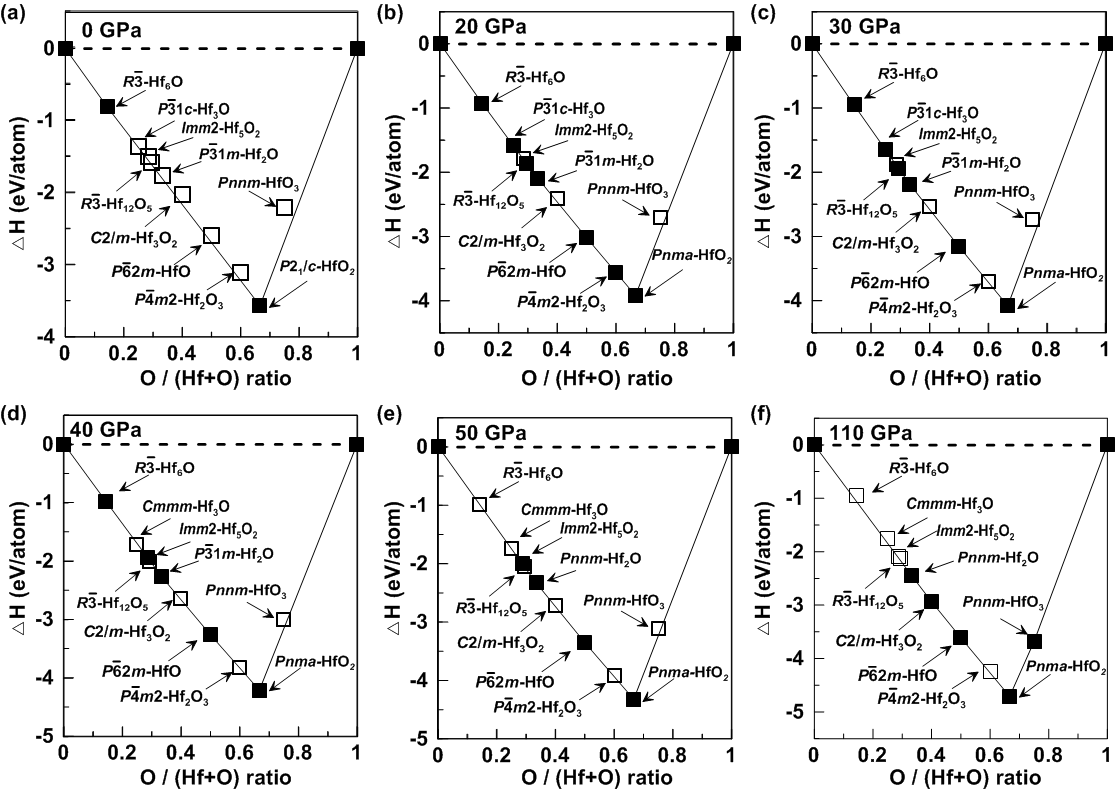}
   \caption{\label{Fig1} Convex hull diagrams for the Hf-O system at (a) 0 GPa, (b) 20 GPa, (c) 30 GPa, (d) 40 GPa, (e) 50 GPa, and (f) 110 GPa, respectively. Solid squares denote stable phases while open squares represent metastable phases.}
     \end{figure}
\end{center}

\begin{center}
\begin{figure}
   \includegraphics[angle=0,width=0.6\linewidth]{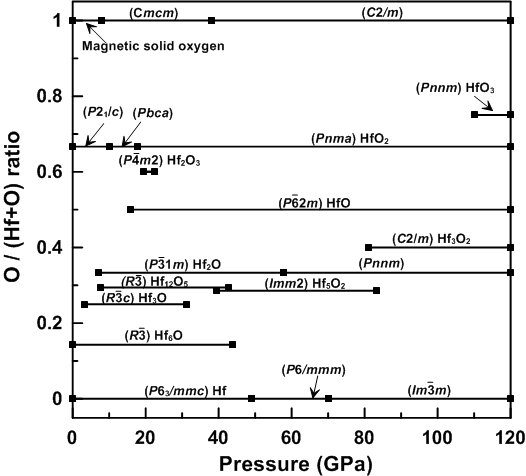}
   \caption{\label{Fig2} Pressure-composition phase diagram of the Hf-O system.}
     \end{figure}
\end{center}

Thermodynamic convex hull, which defines stable compounds, is based on the free energies (at T = 0 K, enthalpies) of the compounds and pure elements in their stable forms. The high-pressure convex hull and pressure-composition phase diagram of the Hf-O system are depicted in Fig. \ref{Fig1} and Fig. \ref{Fig2}, respectively. Besides the three well-known phases of HfO$_2$ and three suboxides (\emph{R}$\bar{3}$-Hf$_6$O, \emph{R}$\bar{3}$\emph{c}-Hf$_3$O and \emph{P}$\bar{3}$1\emph{m}-Hf$_{2}$O), our structure searches found hitherto unknown compounds with new stoichiometries, including Hf$_5$O$_2$, Hf$_3$O$_2$, HfO and HfO$_3$. Note that a new high-pressure phase of Hf$_{2}$O (denoted as \emph{Pnnm}-Hf$_{2}$O) was also found by our searches. Recent work\cite{burton2012first} indicated that Hf$_{12}$O$_5$ is a stable compound at low temperature, disproportionate above 220 K, therefore it is not expected to be observed experimentally. Our work indicates that Hf$_{12}$O$_5$ is actually stable only in the pressure range from 8 GPa to 37 GPa.

Our calculation confirms that Hf$_2$O$_3$ proposed by Xue\cite{xue2013prediction} can exist as a metastable phase (it is dynamically and mechanically stable at 0 GPa), and shows that it should be a stable phase in the pressure range 20-23 GPa. The predicted transition from monoclinic-HfO$_2$ to O\uppercase\expandafter{\romannumeral1}-HfO$_2$ occurs at 10 GPa, which is coincides with experimental observations\cite{desgreniers1999high}. The transition from O\uppercase\expandafter{\romannumeral1}-HfO$_2$ to O\uppercase\expandafter{\romannumeral2}-HfO$_2$ occurs at 18 GPa, which is lower than the experimental result 30-37 GPa\cite{desgreniers1999high,al2010phase} but in good agreement with other theoretical estimates of 17 GPa\cite{al2010phase}. Furthermore, our calculated result shows that OII-HfO$_2$ is stable up to at least 120 GPa, which agrees with previous experimental work\cite{al2010phase}. According to our predictions, only baddeleyite-type HfO$_2$ and \emph{R}$\bar{3}$-Hf$_6$O are stable at 0 GPa, in contrast with the Zr-O system (Zr$_6$O, Zr$_3$O, Zr$_2$O, ZrO and ZrO$_2$ are stable at 0 GPa)\cite{zhang2015novel}.

In order to study the ordering of interstitial oxygen atoms in hcp-HfO$_x$, Hirabayashi et al.\cite{hirabayashi1973superstructure} used electron, neutron and X-ray diffraction to analyze single crystals containing 13.4 at \% O and 15.8 at \% O and found two types of interstitial superstructures: HfO$_{\frac{1}{6}-}$ and HfO$_{\frac{1}{6}+}$ below 600 K. The space group of HfO$_{\frac{1}{6}-}$ reported by Hirabayashi\cite{hirabayashi1973superstructure} is \emph{R}$\bar{3}$, which is identical to our findings. The space group of HfO$_{\frac{1}{6}+}$  is \emph{P}$\bar{3}1$\emph{c} in Hirabayashi' experiment\cite{hirabayashi1973superstructure}. At 0 K and 0 GPa, our results produce three energetically competitive phases for Hf$_{3}$O and their ordering by energy is \emph{R}$\bar{3}$\emph{c}-Hf$_{3}$O (-10.075 eV/atom) $<$ \emph{P}$\bar{3}1$\emph{c}-Hf$_{3}$O (-10.072 eV/atom) $<$ \emph{P}6$_{3}$22-Hf$_{3}$O (-10.069 eV/atom). Therefore, one can note that \emph{P}$\bar{3}1$\emph{c}-Hf$_{3}$O (\emph{P}$\bar{3}1$\emph{c}-Zr$_{3}$O type), exhibits very close but higher energy than \emph{R}$\bar{3}$\emph{c}-Hf$_{3}$O at 0 GPa and 0 K. In order to consider the effects of temperature, quasi-harmonic free-energy of \emph{R}$\bar{3}$\emph{c}-Hf$_{3}$O and \emph{P}$\bar{3}1$\emph{c}-Hf$_{3}$O were calculated using the Phonopy code\cite{togo2008first}. The results indicate that free energy of \emph{P}$\bar{3}1$\emph{c}-Hf$_{3}$O decreases faster than that of \emph{R}$\bar{3}$\emph{c}-Hf$_{3}$O with temperature, enabling \emph{P}$\bar{3}1$\emph{c}-Hf$_{3}$O to become more stable than \emph{R}$\bar{3}$\emph{c}-Hf$_{3}$O at $~$1000 K, thus explaining experimental result.

\begin{center}
\begin{figure}
   \includegraphics[angle=0,width=0.5\linewidth]{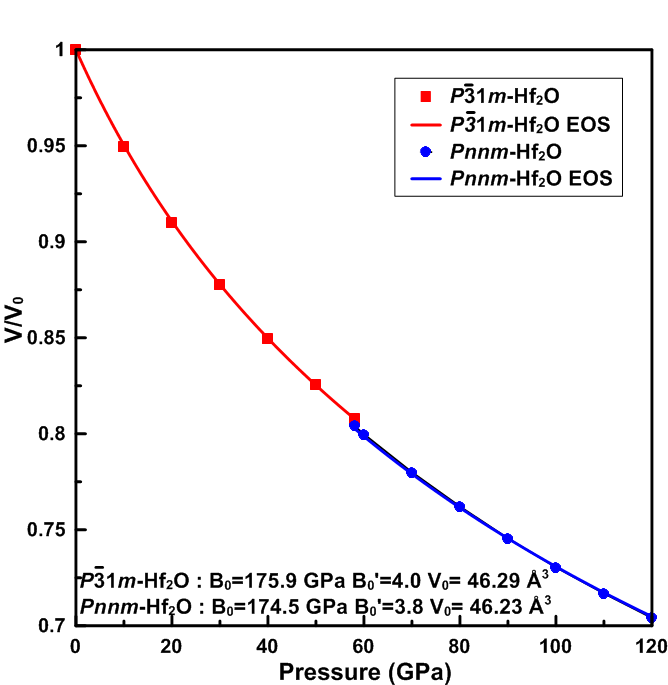}
   \caption{\label{Fig3} (Color online) Equation of state of Hf$_{2}$O. Our calculations were fit to a third-order Birch-Murnaghan equation of state to find B$_{0}$ and B$_{0}^{'}$. }
     \end{figure}
\end{center}

Hf$_{2}$O undergoes a trigonal-to-orthorhombic phase transition at 58 GPa. The crystal structure of the new high-pressure phase \emph{Pnnm}-Hf$_2$O is of anti-CaCl$_2$-type. The Birch-Murnaghan equation of state\cite{birch1952elasticity} was used to fit the compressional behavior of the predicted Hf$_{2}$O phases ( Fig. \ref{Fig3}.). The third-order Birch-Murnaghan EOS is given as

\noindent
\begin{equation}\label{eq1}
P(V)= \frac{3B_{0}}{2}\left [ \left ( \frac{V}{V_{0}} \right )^{-\frac{7}{3}}-\left ( \frac{V}{V_{0}} \right )^{-\frac{5}{3}}\right ]\left \{ 1+\frac{3}{4}(B_{0}^{'}-4)\left [ \left ( \frac{V}{V_{0}} \right )^{-\frac{2}{3}}-1 \right ] \right \}
\end{equation}

Three parameters are used to describe the EOS: the volume at 0 GPa (V$_{0}$), the bulk modulus at 0 GPa (B$_{0}$), and the first pressure derivative of the the bulk modulus at 0 GPa (B$_{0}^{'}$). Most materials have 3 $\leq$ B$_{0}^{'}$ $\leq$ 6\cite{jeanloz1988universal,birch1978finite}. The B$_{0}^{'}$ of \emph{P}$\bar{3}$1\emph{m}-Hf$_2$O and \emph{Pnnm}-Hf$_2$O is 4.0 and 3.8, respectively.

\subsection{Structure character in Hf-O compounds}

\begin{center}
\begin{figure}
   \includegraphics[angle=0,width=1.0\linewidth]{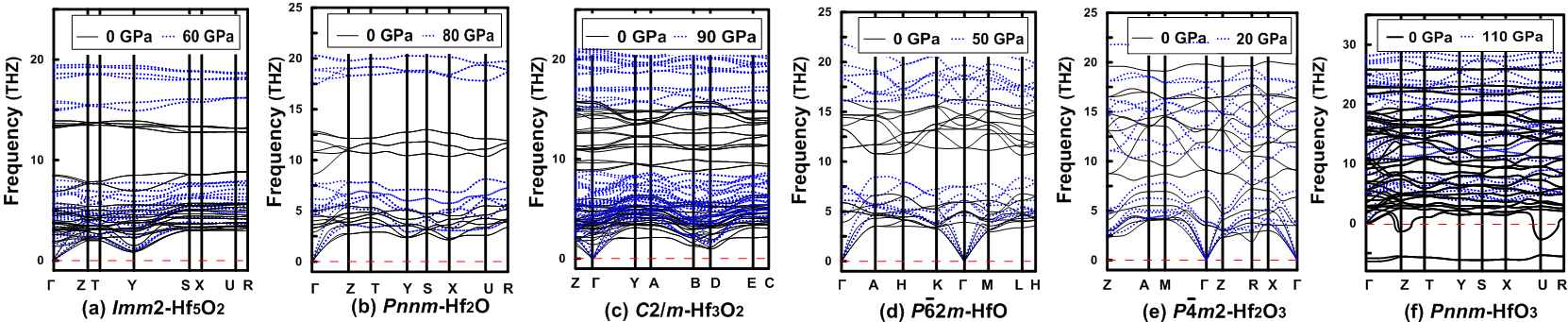}
   \caption{\label{Fig4} (Color online) Calculated phonon dispersion curves for the  (a) \emph{Imm}2-Hf$_5$O$_2$ at 0 and 60 GPa  (b) \emph{Pnnm}-Hf$_2$O at 0 and 80 GPa (c) \emph{C}2/\emph{m}-Hf$_3$O$_2$ at 0 and 90 GPa (d) \emph{P}$\bar{6}$2\emph{m}-HfO at 0 and 50 GPa (e) \emph{P}$\bar{4}$\emph{m}2-Hf$_2$O$_3$ at 0 and 20 GPa (f) \emph{Pnnm}-HfO$_3$ at 0 and 110 GPa. The solid black and dashed blue lines represent the results at zero and high pressures, respectively.}
\end{figure}
\end{center}

\begin{center}
\begin{figure}
   \includegraphics[angle=0,width=0.5\linewidth]{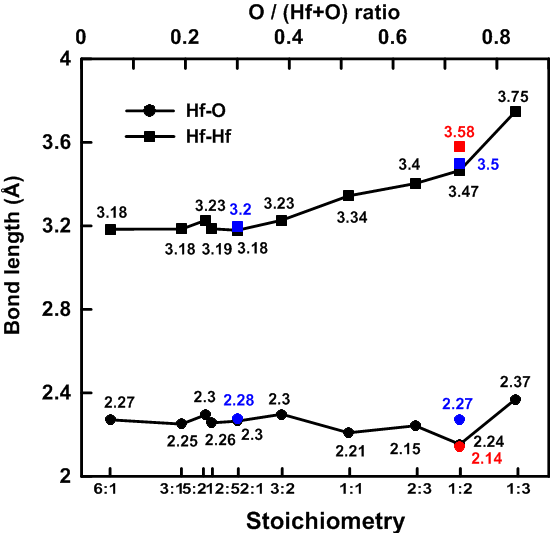}
   \caption{\label{Fig5} (Color online) Average bond lengths in Hf-O compounds at 0 GPa.}
     \end{figure}
\end{center}

Table \ref{Tab1} lists the detailed crystallographic data of \emph{Imm}2- Hf$_5$O$_2$, \emph{Pnnm}- Hf$_2$O, \emph{C}2/\emph{m}-Hf$_3$O$_2$, \emph{P}$\bar{6}$2\emph{m}-HfO, \emph{C}2/\emph{m}-Hf$_2$O$_3$ and \emph{Pnnm}-HfO$_3$ compounds at 0 GPa. The dynamical stabilities of all the new phases are checked by calculating phonon dispersion. As shown in Fig.\ref{Fig4}, except for HfO$_{3}$, no imaginary phonon frequencies are found in the whole Brillouin zone at both ambient and high pressure, which means that they are dynamically stable and probably quenchable to ambient pressure. In contrast, HfO$_{3}$ is stable only at high pressure, but at 0 GPa shows total dynamical instability and most likely decomposes. The special electronic structure of HfO$_3$ will be discussed below. The weighted average lengths of Hf-Hf and Hf-O bonds in Hf-O compounds are plotted in Fig. \ref{Fig5}.

\begin{table}
\centering
\caption{Structural parameters of \emph{Imm}2- Hf$_5$O$_2$, \emph{Pnnm}- Hf$_2$O, \emph{C}2/\emph{m}-Hf$_3$O$_2$, \emph{P}$\bar{6}$2\emph{m}-HfO, \emph{C}2/\emph{m}-Hf$_2$O$_3$ and \emph{Pnnm}-HfO$_3$ at 0 GPa.}
\begin{tabular}{ c c  c c c  c c c }   
\hline \hline
Compound   & Space group   & Enthalpy of formation   & Lattice constants & Wyckoff positions  & x&y &z \\
           &               & (eV/atom)       &(\AA)             &                &   &  &  \\
\hline
Hf$_5$O$_2$ & \emph{Imm}2               & -1.52 & $a$=14.455     & Hf 4c  & 0.711 & 0.50  & 0.566 \\
            &                           &       & $c$=3.141      & Hf 2b  & 0.00  & 0.50  & 0.098  \\
            &                           &       &               & Hf 4c  & 0.097 & 0.00  & 0.594 \\
            &                           &       & $c$=5.082      & O 4c   & 0.645 & 0.00  & 0.818 \\
Hf$_2$O     & \emph{Pnnm}               & -1.76 & $a$=5.092      & Hf 4g  & 0.263 & 0.341 & 0.50\\
            &                           &       & $b$=5.723      & O 2c   & 0.00  & 0.50  & 0.00 \\
            &                           &       & $c$=3.175      &        &       &       &  \\
Hf$_3$O$_2$ & \emph{C}2/\emph{m}        & -2.04 & $a$=11.967     & Hf 4i  & 0.625 & 0.50  & 0.007\\
            &                           &       & $b$=3.131      & Hf 4i  & 0.465 & 0.00  & 0.346 \\
            &                           &       & $c$=11.198     & Hf 4i  & 0.286 & 0.00  & 0.676  \\
            &                           &       & $\beta=99.67^\circ$& O 4i& 0.378& 0.50 & 0.607  \\
            &                           &       &               & O 4i   &  0.787& 0.00  & 0.192   \\
HfO         & \emph{P}$\bar{6}$2\emph{m}& -2.60 & $a$=5.230      & Hf 1b  & 0.00  & 0.00  & 0.50\\
            &                           &       & $c$=3.187      & Hf 2c  & 0.667 & 0.333 & 0.00 \\
            &                           &       &               & O 3g   & 0.00  & 0.592 & 0.50 \\
Hf$_2$O$_3$ & \emph{P}$\bar{4}$\emph{m}2& -3.11 & $a$=3.137      & Hf 2g  & 0.00  & 0.50  & 0.744\\
            &                           &       & $c$=5.638      & O  2g  & 0.00  & 0.50  & 0.135 \\
            &                           &       &               & O 1c   & 0.50  & 0.50  & 0.50 \\
HfO$_3$      &\emph{Pnnm}                & -2.22 & $a$=5.554     & Hf 4c  & 0.246 & 0.110 & 0.250\\
             &                           &       & $b$=6.457     & O 4c   & 0.359 & 0.426 & 0.250 \\
             &                           &       & $c$=3.307     & O 4c   & 0.025 & 0.339 &0.750  \\

\hline\hline
\end{tabular}
\label{Tab1}
\end{table}

\begin{center}
\begin{figure}
   \includegraphics[angle=0,width=0.5\linewidth]{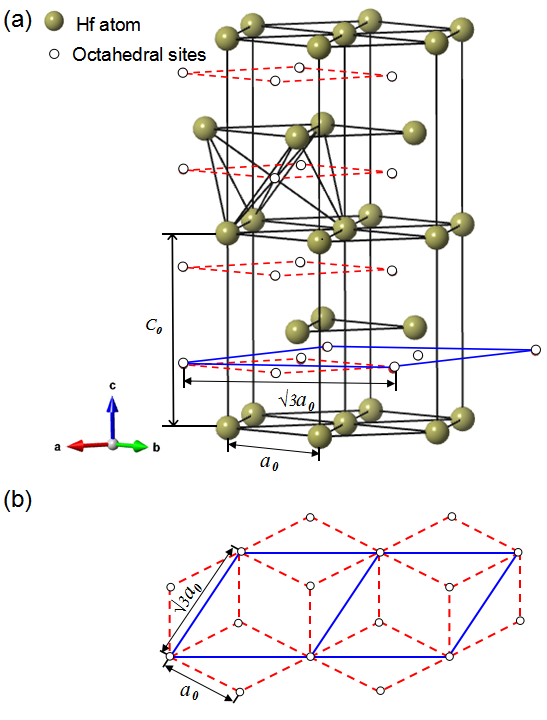}
   \caption{\label{Fig6} (Color online) Octahedral voids in hcp hafnium.}
     \end{figure}
\end{center}

\begin{center}
\begin{figure}
   \includegraphics[angle=0,width=1.0\linewidth]{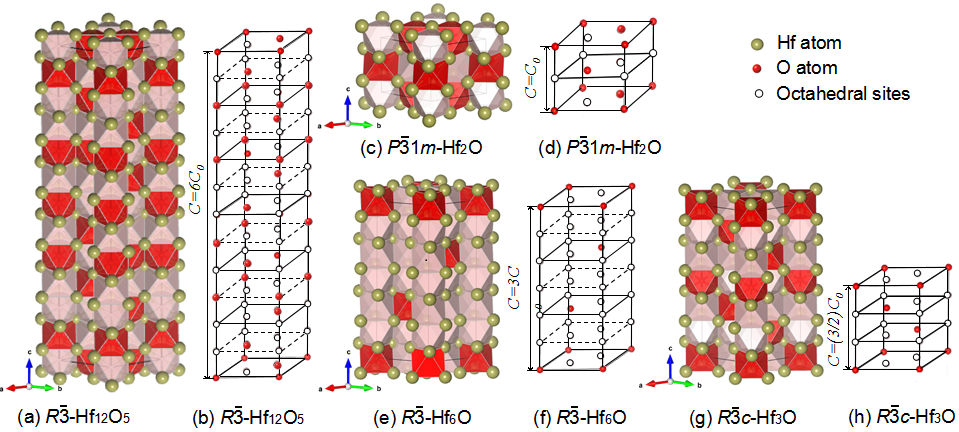}
   \caption{\label{Fig7} (Color online) Oxygen sublattice representation (arrangement of oxygen atoms in the octahedral interstitial sites) and polyhedral representation of (a)\&(b) \emph{R}$\bar{3}$-Hf$_{12}$O$_5$ (c)\&(d) \emph{P}$\bar{3}$1\emph{m}-Hf$_2$O (e)\&(f) \emph{R}$\bar{3}$-Hf$_6$O (g)\&(h) \emph{R}$\bar{3}$\emph{c}-Hf$_3$O. Oxygen-centered octahedra and oxygen vacancies are shown in red and pink polyhedra, respectively. Oxygen sublattice representations (b, d, f, h) show only oxygen atoms (filled circles) and vacancies (open circles).}
     \end{figure}
\end{center}

Structurally, hafnium oxides can be divided into four groups: suboxides with oxygen interstitials in hcp-Hf (Hf$_6$O, Hf$_3$O, Hf$_{12}$O$_5$ and \emph{P}$\bar{3}$1\emph{m}-Hf$_2$O); other suboxides (Hf$_5$O$_2$, \emph{Pnnm}-Hf$_2$O and HfO); normal oxides (Hf$_2$O$_3$, HfO$_2$) and oxide peroxide (HfO$_3$). The octahedral sites of hcp hafnium metal are depicted in Fig.\ref{Fig6}. Oxygen atoms prefer to occupy these octahedral sites and form ordered structures \emph{R}$\bar{3}$-Hf$_6$O, \emph{R}$\bar{3}$\emph{c}-Hf$_3$O, \emph{R}$\bar{3}$-Hf$_{12}$O$_5$ and \emph{P}$\bar{3}$1\emph{m}-Hf$_2$O, as shown in Fig. \ref{Fig7} (a) (b) (c) and (d), where Hf atom sites are omitted. The polyhedral representation of these structures is shown in Fig. \ref{Fig6} (e) (f) (g) and (h). Anti-CaCl$_2$-type (\emph{Pnnm}) structure of Hf$_2$O can also be represented as an hcp-sublattice (distorted) of Hf atoms, where half of octahedral voids are occupied by O atoms. The structure of Hf$_3$O$_2$ can be considered to be defective because each layer lacks some Hf atoms to form a Hf-graphene layer. These vacancies are responsible for low values of the mechanical properties of Hf$_3$O$_2$.

\begin{center}
\begin{figure}
   \includegraphics[angle=0,width=1.0\linewidth]{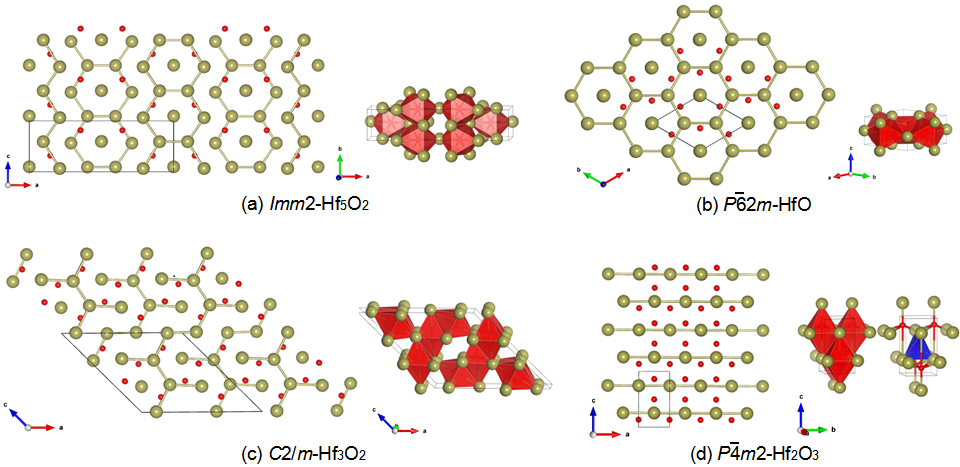}
   \caption{\label{Fig8} (Color online) Crystal structures of (a) \emph{Imm}2-Hf$_5$O$_2$ (b) \emph{P}$\bar{6}$2\emph{m}-HfO (c) \emph{P}$\bar{4}$\emph{m}2-Hf$_2$O$_3$ and (d) \emph{P}$\bar{4}$\emph{m}2-Hf$_2$O$_3$. O-centered octahedra and O-centered tetrahedra are shown in red and blue polyhedra, respectively. Large spheres-Hf atoms; small spheres-O atoms.}
     \end{figure}
\end{center}

Similar with ZrO, the structure of HfO contains Hf-graphene layers stacked on top of each other (Zr-Zr distances within the layer are 3.01 \AA, and between the layers 3.18 \AA), as illustrated in Fig. 8(b), as well as additional Hf and O atoms. The structure can be represented as $\omega$-phase of Hf, intercalated with oxygen atoms. This structure, therefore, is built by a 3D-framework of short and strong Hf-O bonds, reinforced by rather strong Hf-Hf bonds. The former lead to high hardness, the latter may improve toughness due to semimetallic behavior. \emph{P}$\bar{4}$\emph{m}2-Hf$_2$O$_3$, which was firstly proposed by Xue\cite{xue2013prediction}, has 8-fold and 6-fold coordination of Hf atoms, as shown in Fig \ref{Fig8}.

\begin{center}
\begin{figure}
   \includegraphics[angle=0,width=0.4\linewidth]{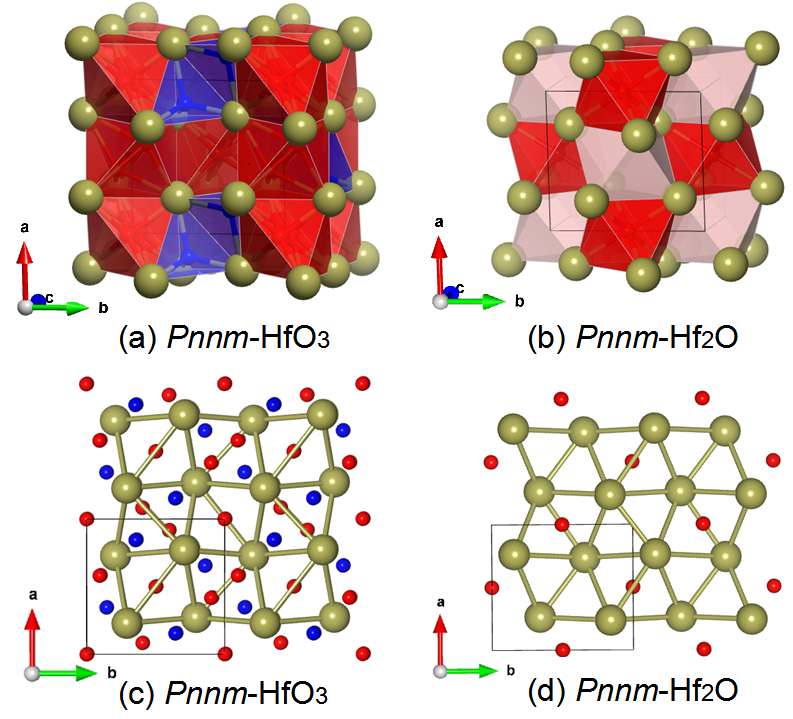}
   \caption{\label{Fig9} (Color online) Crystal structure of (a) \emph{Pnnm}-HfO$_3$ (b) \emph{Pnnm}-Hf$_2$O; (c) \emph{Pnnm}-HfO$_3$ (d) \emph{Pnnm}-Hf$_2$O.}
     \end{figure}
\end{center}

\emph{Pnnm}-HfO$_3$ becomes stable at pressures above to 110 GPa. This high-pressure phase originally derives from oxygen atom dissolving in both octahedral and tetrahedral voids of a heavily distorted hcp-Hf, as shown in Fig 9(a).
However, due to short distances between tetrahedral voids in the hcp structures, some O atoms form pairs and as a result HfO$_3$ simultaneously contains oxide O$^{2-}$ and peroxide [O-O]$^{2-}$ anions, and can be described as "oxide peroxide". The O-O bond length in HfO$_3$ is 1.44 \AA  at 110 GPa, which is a little smaller than the O-O bond length in peroxide [O-O]$^{2-}$ ion with 1.47 \AA\cite{wells1986} at ambient conditions. It seems that peroxides and oxide peroxides (e.g. Al$_4$O$_7$ and AlO$_2$) become stabilized in many systems under pressure\cite{liu2015prediction}.

\subsection{Mechanical properties of Hf-O compounds}

\begin{center}
\begin{figure}
   \includegraphics[angle=0,width=0.5\linewidth]{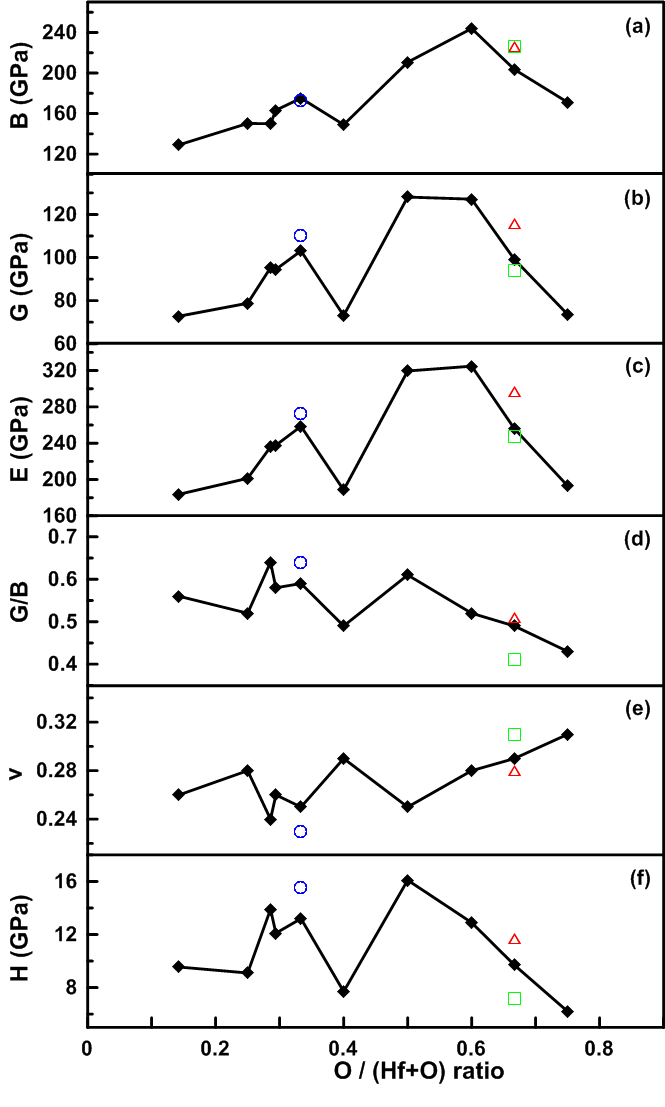}
   \caption{\label{Fig10} (Color online) Compositional dependence of the computed mechanical properties of Hf-O compounds. The blue open circle represents \emph{Pnnm}-Hf$_2$O; red open triangle represents O\uppercase\expandafter{\romannumeral1}-HfO$_2$ ; green open square represents O\uppercase\expandafter{\romannumeral2}-HfO$_2$.}
     \end{figure}
\end{center}

Previous studies\cite{ohtaka2001phase,haines1997characterization,desgreniers1999high} suggested that dense high-pressure phase OII-HfO$_2$ is quenchable to ambient conditions and has a high bulk modulus, and might be superhard (H $>$ 40 GPa). However recent study\cite{al2010phase} reported that the hardness of OII-HfO$_2$ is well below 40 GPa and therefore this phase is not superhard. Interestingly, our systematic results not only confirm known hardness of HfO$_2$ polymorphs: H(OII) $<$ H(MI) $<$ H(OI), but also suggest that HfO has the highest hardness among all hafnium oxides, see Fig. \ref{Fig10}(f). In addition, \emph{Pnnm}-Hf$_2$O and \emph{Imm}2-Hf$_5$O$_2$ also exhibit higher hardness than other Hf-O compounds, as shown in Tab \ref{Tab2}. The hardness of Hf-O compounds does not monotonically change with O content, but a maximum at HfO.

The calculated modulus $\emph{B}$, shear modulus $\emph{G}$, Young's modulus $\emph{E}$, Poisson's ratio $\upsilon$, and hardness of all stable Hf$-$O compounds are depicted in Table\ref{Tab2} and Fig.\ref{Fig10} (for comparison, the elastic data of the high-pressure phase \emph{Pnnm}-HfO$_3$ are reported at 0 GPa although it is unstable at 0 GPa.) From Fig.\ref{Fig10} we can conclude that the high O content in the crystal does not guarantee high hardness of Hf-O compounds and the structure plays an important role in determining mechanical properties as we discussed above. The Vickers hardness was calculated according to Chen's model\cite{chen2011modeling}:

\begin{equation}\label{eq4}
H_V = 2 * (k^2*G)^{0.585}-3
\end{equation}

\begin{table}
\centering
\caption{Calculated bulk modulus $\emph{B}$, shear modulus $\emph{G}$, Young's modulus $\emph{E}$, Poisson's ratio $\upsilon$ and hardness of Hf-O compounds, compared with literature data for HfO$_2$ at 0 GPa. All properties are in GPa (except dimensionless G/B and $\upsilon$ ).}
\begin{tabular}{ c c c  c c c  c c c  c}   
\hline \hline
Compound    &           & Space group               & P & B$_H$  & G$_H$ & E  &  G/B  & $\upsilon$ & H$_v$ \\
\hline
Hf$_6$O     & This work & \emph{R}$\bar{3}$         & 0 & 129.2 & 72.8  & 183.8 & 0.56 & 0.26 & 9.55 \\
Hf$_3$O     & This work & \emph{R}$\bar{3}$\emph{c} & 0 & 150.3 & 78.8  & 201.3 & 0.52 & 0.28 & 9.1\\
Hf$_5$O$_2$ & This work & \emph{Imm}2               & 0 & 150.0 & 95.3  & 235.9 & 0.64 & 0.24 & 13.9 \\
Hf$_{12}$O$_5$& This work & \emph{R}$\bar{3}$         & 0 & 163.3 & 94.5  & 237.7 & 0.58 & 0.26 & 12.1 \\
Hf$_2$O     & This work & \emph{P}$\bar{3}$1\emph{m}& 0 & 175.2 & 103.1 & 258.6 & 0.59 & 0.25 &13.2\\
Hf$_2$O     & This work & \emph{Pnnm}               & 0 & 173.0 & 110.3 & 272.9 & 0.64 & 0.23 & 15.5\\
Hf$_3$O$_2$ & This work & \emph{C}2/\emph{m}        & 0 & 154.2 & 75.9  & 195.6 & 0.49 & 0.29 & 8.0\\
HfO         & This work & \emph{P}$\bar{6}$2\emph{m}& 0 & 210.7 & 128.1 & 319.5 & 0.61 & 0.25 & 16.1\\
Hf$_2$O$_3$ & This work & \emph{P}$\bar{4}$\emph{m}2& 0 & 243.9 & 127.1 & 324.8 & 0.52 & 0.28 & 12.9\\
HfO$_2$     & This work &\emph{P}2$_1$/\emph{c}     & 0 & 203.6 & 99.2  & 256.1 & 0.49 & 0.29 & 9.7\\
            & Experiment\cite{okutomi1984sintering}&& 0 &     &       &       &      &      & 9.9\\
HfO$_2$     & This work &\emph{Pbca}                & 0 & 225.9 & 115.8 & 296.6 & 0.51 & 0.28 & 11.7\\
HfO$_2$     & This work &\emph{Pnma}                & 0 & 226.3 & 93.8  & 247.3 & 0.41 & 0.31 & 7.2\\
            & Experiment\cite{haines1997proceeding}&& 0 &      &       &       &      &      & 6-13\\
HfO$_3$     & This work &\emph{Pnnm}                & 0 &171.1 & 73.6  & 193.0 & 0.43 & 0.31 & 6.2\\
\hline\hline
\end{tabular}
\label{Tab2}
\end{table}

\begin{center}
\begin{figure}
   \includegraphics[angle=0,width=1.0\linewidth]{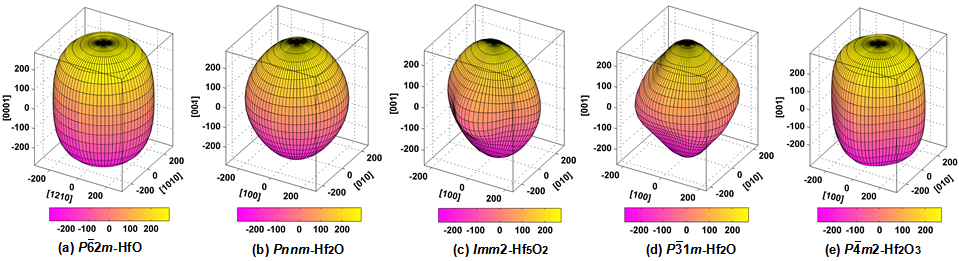}
   \caption{\label{Fig11} (Color online) Orientational dependence of Young's moduli (in GPa) of (a) \emph{P}$\bar{6}$2\emph{m}-HfO (b) \emph{Pnnm}-Hf$_2$O (c) \emph{Imm}2-Hf$_5$O$_2$  (d) \emph{P}$\bar{3}$1\emph{m}-Hf$_2$O and (e) \emph{P}$\bar{4}$\emph{m}2-Hf$_2$O$_3$.}
     \end{figure}
\end{center}

We calculated the elastic anisotropy of five special phases: \emph{P}$\bar{6}$2\emph{m}-HfO, \emph{Pnnm}-Hf$_2$O, \emph{Imm}2-Hf$_5$O$_2$, \emph{P}$\bar{3}$1\emph{m}-Hf$_{2}$O and \emph{P}$\bar{4}$\emph{m}2-Hf$_2$O$_3$. As shown in Fig. \ref{Fig11}, all of these five phases exhibit a moderate amount of anisotropy of Young's modulus. The directional dependence of the Young's modulus for hexagonal, orthorhombic, trigonal and tetragonal crystals can be calculated as:

\noindent
\begin{equation}\label{eq2}
\frac{1}{E_{hex}} = {s_{11}}{(1 - l_3^2)^2} + {s_{33}}l_3^4 + (2{s_{13}} + {s_{44}})l_3^2(1 - l_3^2)
\end{equation}

\noindent
\begin{equation}\label{eq3}
\frac{1}{E_{ortho}} = s_{11}(l_{1})^{4}+s_{22}(l_{2})^{4}+s_{33}(l_{3})^{4}+l_{2}^{2}l_{3}^{2}(2s_{23}+s_{44})+l_{1}^{2}l_{3}^{2}(2s_{13}+s_{55})+l_{2}^{2}l_{1}^{2}(2s_{12}+s_{66})
\end{equation}

\noindent
\begin{equation}\label{eq4}
\frac{1}{E_{tri}} = (1 - l_3^2){s_{11}} + l_3^4{s_{33}} + l_3^2(1 - l_3^2)(2{s_{13}} + {s_{44}}) + 2{l_2}{l_3}(3l_1^2 - l_2^2){s_{14}},
\end{equation}

\noindent
\begin{equation}\label{eq5}
\frac{1}{E_{tetra}} = s_{11}(l_{1}^{4}+l_{2}^{4})+s_{33}l_{3}^{4}+(2s_{12}+s_{66})l_{1}^{2}l_{2}^{2}+(2s_{13}+s_{44})l_{3}^{2}(1-l_{3}^{2})
\end{equation}
\noindent
where $s_{11}$, $s_{12}$, etc., are the elastic compliance constants and $l_1$, $l_2$, $l_3$ are the direction cosines of a particular crystallographic orientation to coordinate axes $x_1$, $x_2$ and $x_3$, respectively.

\subsection{Electronic structure of Hf-O compounds}

Fig. \ref{Fig13} shows band structures of Hf-O compounds at 0 GPa (including phases stable at both zero and high pressure). Total and partial densities of states (DOS) are presented in Fig. \ref{Fig14}. \emph{R}$\bar{3}$-Hf$_6$O, \emph{R}$\bar{3}$\emph{c}-Hf$_3$O, \emph{Imm}2- Hf$_5$O$_2$, \emph{R}$\bar{3}$- Hf$_{12}$O$_5$, \emph{P}$\bar{3}$1\emph{m}-Hf$_2$O, \emph{Pnnm}- Hf$_2$O and \emph{C}2/\emph{m}-Hf$_3$O$_2$ are predicted to be metallic with a sizable density of states at the Fermi level. The DOSs of \emph{R}$\bar{3}$-Hf$_6$O, \emph{R}$\bar{3}$\emph{c}-Hf$_3$O, \emph{Imm}2- Hf$_5$O$_2$, \emph{R}$\bar{3}$- Hf$_{12}$O$_5$, \emph{P}$\bar{3}$1\emph{m}-Hf$_2$O, \emph{Pnnm}- Hf$_2$O and \emph{C}2/\emph{m}-Hf$_3$O$_2$ below E$_{F}$ are mainly due to Hf-d and O-p orbitals, and the interactions between the Hf-d orbitals are responsible for metallicity.

\begin{center}
\begin{figure}[h!]
   \includegraphics[angle=0,width=0.35\linewidth]{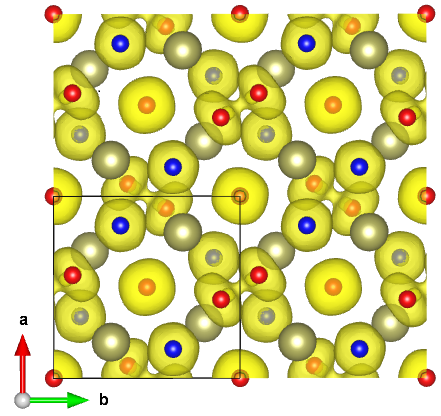}
   \caption{\label{Fig12} (Color online) ELF isosurface (ELF = 0.62) for HfO$_3$. Blue and red atoms represent oxide O$^{2-}$ and peroxide [O-O]$^{2-}$ ions, respectively. }
     \end{figure}
\end{center}

\begin{center}
\begin{figure}
   \includegraphics[angle=0,width=1.0\linewidth]{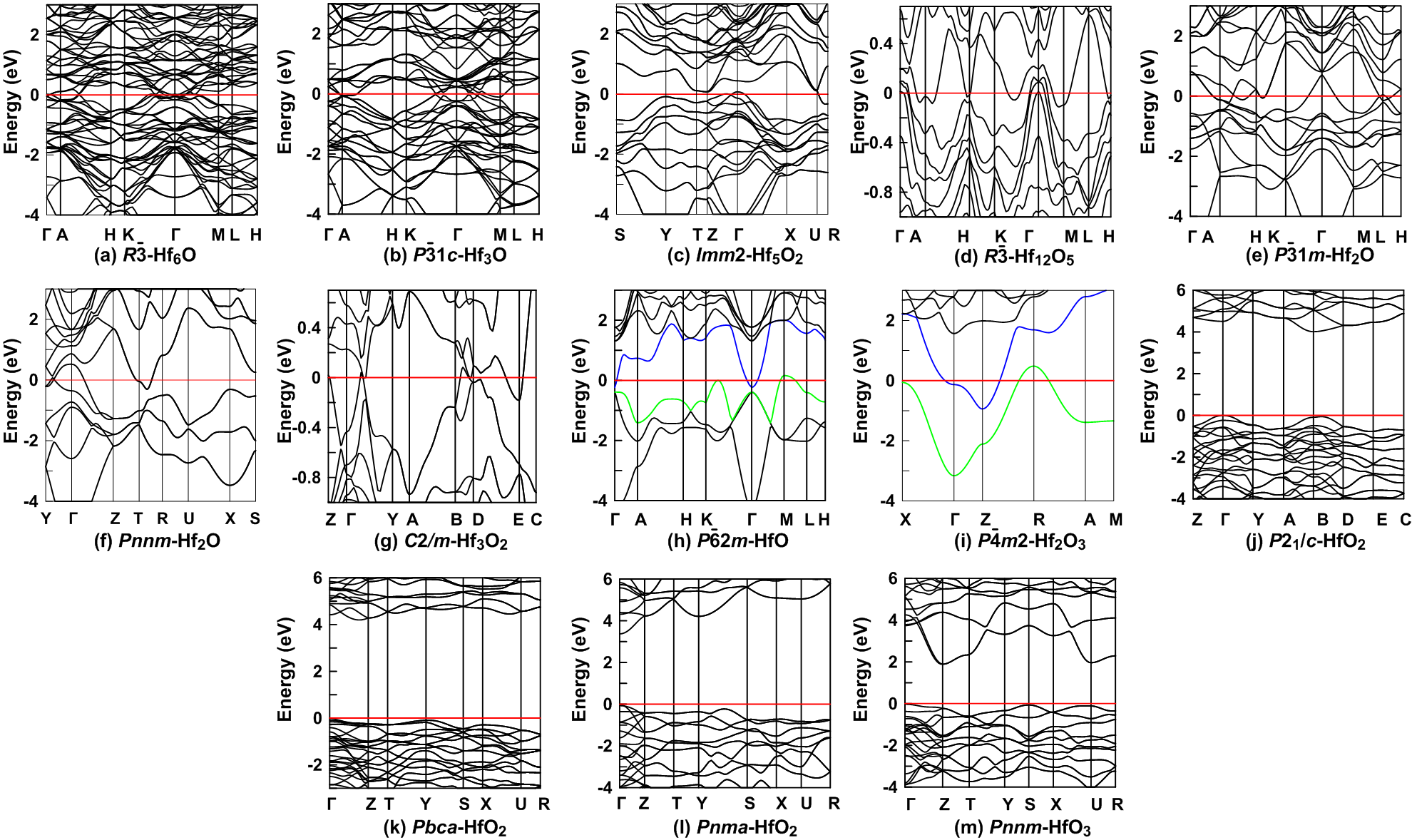}
   \caption{\label{Fig13} (Color online) Band structures of hafnium oxides at 0 GPa. The Fermi energy is set to zero.}
     \end{figure}
\end{center}

\begin{center}
\begin{figure}
   \includegraphics[angle=0,width=0.5\linewidth]{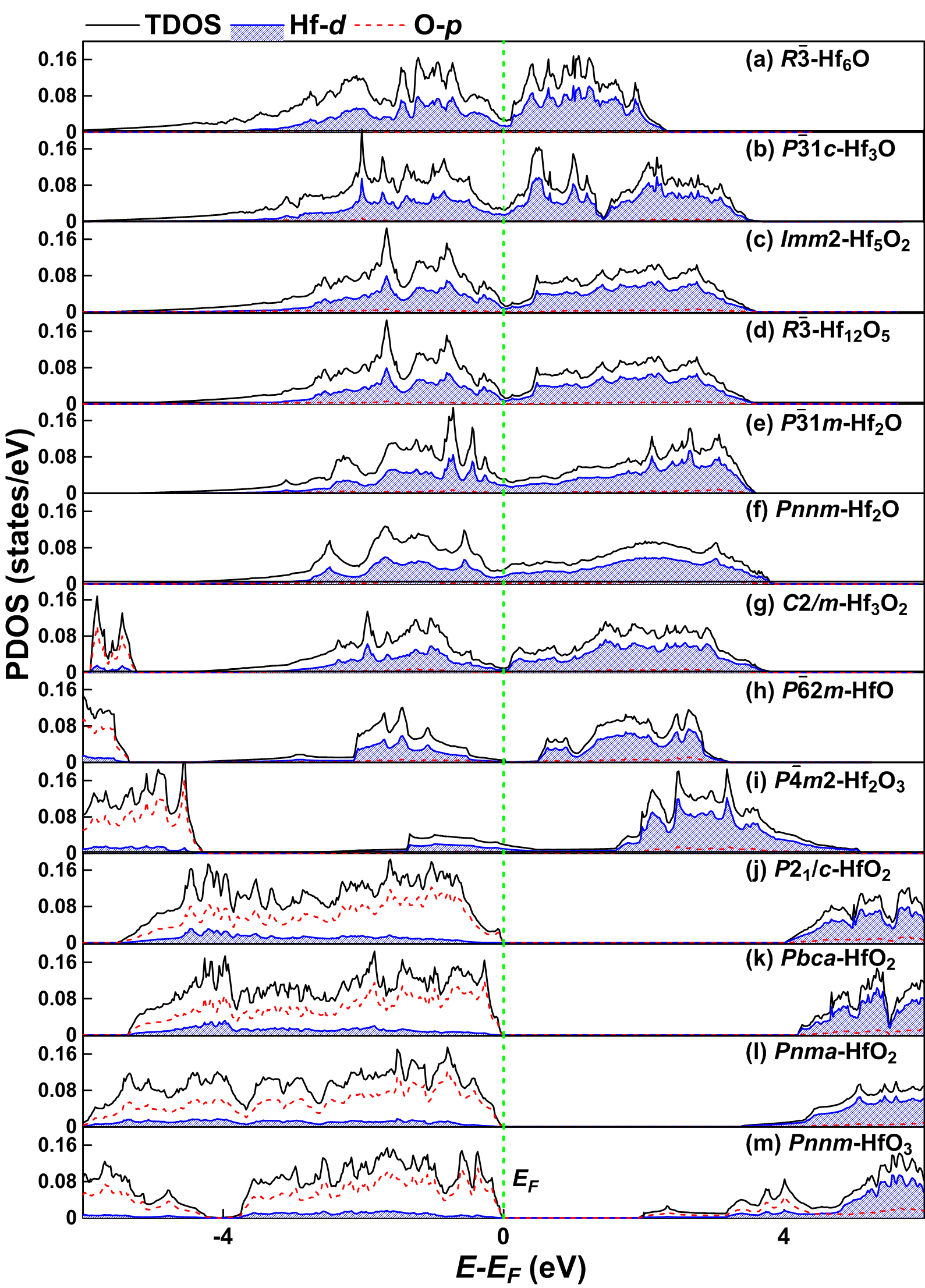}
   \caption{\label{Fig14} (Color online)  The normalized (per electron) total (TDOS) and partial densities of states (PDOS) of hafnium oxides at 0 GPa. The Fermi energy is set to zero.}
     \end{figure}
\end{center}

Unlike regular metals, semimetals possess both electronic and hole conduction, which can be seen in the band structure as overlap of the partially vacant valence band top and occupied conduction band bottom located at different points in the Brillouin Zone\cite{xue2013prediction}. Moreover, the upper limit of electron and hole density for a semimetal should below 10$^{22}$ cm$^{-3}$. The carrier concentrations of the most common semimetals, for example Bi, Sb and As are 3$\times$10$^{17}$ cm$^{-3}$, 5$\times$10$^{19}$ cm$^{-3}$, and 2$\times$10$^{20}$ cm$^{-3}$\cite{liu1995electronic,ashcroft1976solid}, respectively. Semimetallic behavior of \emph{P}$\bar{6}$2\emph{m}-HfO and \emph{P}$\bar{4}$\emph{m}2-Hf$_2$O$_3$ can be reflected in the calculated band structures, as shown in Fig. \ref{Fig12}(h,i) and in the DOS diagrams (Fig. \ref{Fig14}(h,i)), showing very few states at the Fermi level. For the band structure of \emph{P}$\bar{6}$2\emph{m}-HfO, there are small electron and hole pockets at $\Gamma$ and M, respectively, but there are no band crossings between the lowest unoccupied bands and the highest occupied bands. Both band edges are mostly derived from Hf 5d states. In the case of \emph{P}$\bar{4}$\emph{m}2-Hf$_2$O$_3$, partially occupied valence band top and conduction band bottom correspond to different high symmetry points R and Z, respectively. The electron and hole densities can be estimated by integrating the occupation numbers in the 3D Brillouin zone. The electron and hole densities of HfO are both 1.1$\times$10$^{20}$ cm$^{-3}$ by integrating their occupation of the blue and green bands shown in Fig. \ref{Fig13}(h), respectively. For \emph{P}$\bar{4}$\emph{m}2-Hf$_2$O$_3$, the electron and hole densities are both 2.1$\times$10$^{21}$ cm$^{-3}$, which is very close to 1.8$\times$10$^{21}$ cm$^{-3}$ obtained in Xue's work\cite{xue2013prediction}.

The DFT band gaps of \emph{P}2$_1$/\emph{c}-HfO$_2$, \emph{Pbca}-HfO$_2$, \emph{Pnma}-HfO$_2$ and \emph{Pnnm}-HfO$_3$ are 4.01 eV, 4.18 eV, 3.36 eV and 1.92 eV, respectively, and the highest occupied states are all derived mainly from O-p orbitals, as shown in Fig. \ref{Fig14}(j, k, l and m). Therefore, according to their electronic character, Hf-O compounds can be divided into three types: metallic, including \emph{R}$\bar{3}$-Hf$_6$O, \emph{R}$\bar{3}$\emph{c}-Hf$_3$O, \emph{Imm}2- Hf$_5$O$_2$, \emph{R}$\bar{3}$- Hf$_{12}$O$_5$, \emph{P}$\bar{3}$1\emph{m}-Hf$_2$O, \emph{Pnnm}- Hf$_2$O and \emph{C}2/\emph{m}-Hf$_3$O$_2$; semimetallic, including \emph{P}$\bar{6}$2\emph{m}-HfO and \emph{P}$\bar{4}$\emph{m}2-Hf$_2$O$_3$; insulating or semiconducting, including \emph{P}2$_1$/\emph{c}-HfO$_2$, \emph{Pbca}-HfO$_2$, \emph{Pnma}-HfO$_2$ and \emph{Pnnm}-HfO$_3$. Electron localization function (ELF) clearly reveals special featrure of HfO$_3$, the coexistence of oxide O$^{2-}$ and peroxide [O-O]$^{2-}$ anions (Fig. \ref{Fig12}). The peroxide is responsible for gap states, which significantly reduce the electronic band gap of HfO$_2$ (Fig. \ref{Fig14} (m)). To obtain further insight, we applied the Atoms in Molecules (AIM) theory developed by Bader \cite{bader1990atoms}. Bader charges are +2.5 for Hf, -0.68 for peroxide anion and -1.16 for oxide anion in HfO$_3$ at 110 GPa, which shows a significantly ionic character of bonding.

\section{Conclusions}

We have systematically predicted stable compounds and crystal structures in the Hf-O system at pressures up to 120 GPa using ab initio evolutionary algorithm USPEX. Several new stable compounds, including \emph{Imm}2-Hf$_5$O$_2$, \emph{C}2/\emph{m}-Hf$_3$O$_2$, \emph{P}$\bar{6}$2\emph{m}-HfO and \emph{Pnnm}-HfO$_3$ are found for the first time. \emph{Pnnm}-Hf$_2$O, which is the new high-pressure phase of Hf$_2$O, is also discovered. HfO$_3$ shows interesting structure, simultaneously containing oxide O$^{2-}$ and peroxide [O-O]$^{2-}$ anions. Semimetallic properties of \emph{P}$\bar{6}$2\emph{m}-HfO and \emph{P}$\bar{4}$\emph{m}2-Hf$_2$O$_3$ are demonstrated through their band structures, as well as low densities of conduction electrons and holes. Our results demonstrate that Hf$_{3}$O$_{2}$ is more ductile than other Hf-O compounds, and the hardest compound is HfO instead of O\uppercase\expandafter{\romannumeral2}-HfO$_2$ . The superior mechanical properties of \emph{P}$\bar{6}$2\emph{m}-HfO, such as bulk modulus $\emph{B}$, shear modulus $\emph{G}$, Young's modulus $\emph{E}$ and hardness \emph{H$_{v}$}, can be attributed to the peculiar combination of strong Hf-O and Hf-Hf bonds. \emph{Pnnm}-Hf$_2$O, \emph{Imm}2-Hf$_5$O$_2$, \emph{P}$\bar{3}$1\emph{m}-Hf$_{2}$O and \emph{P}$\bar{4}$\emph{m}2-Hf$_2$O$_3$ also show excellent mechanical properties. Clearly, high O content is not a key factor affecting the mechanical properties of Hf-O compounds. Suboxides: Hf$_6$O, Hf$_3$O, Hf$_12$O$_5$ and \emph{P}$\bar{3}$1\emph{m}-Hf$_2$O based on hcp-Hf sublattice provide easy pathways for absorbing or desorbing oxygen. The recognition of the common structural features between \emph{P}$\bar{6}$2\emph{m}-HfO and $\omega$-Hf gives further insight into the physical properties and suggests that HfO can be made as a hard semimetallic coating on $\omega$-Hf substrate. \emph{Pnnm}-Hf$_2$O, \emph{Imm}2-Hf$_5$O$_2$, \emph{P}$\bar{3}$1\emph{m}-Hf$_{2}$O and \emph{P}$\bar{4}$\emph{m}2-Hf$_2$O$_3$ phases in particular can be quenched to ambient pressure and can be candidates for applications requiring mechanically strong materials.

\section{acknowledgments}
This work was supported by the National Science Foundation (EAR-1114313), DARPA (Grants No. W31P4Q1210008), the Basic Research Foundation of NWPU (No. JCY20130114), the Natural Science Foundation of China (No. 51372203, 51332004), the Foreign Talents Introduction, the Academic Exchange Program of China (No. B08040) and the Government (No. 14.A12.31.0003) of Russian Federation. The computational resources at High Performance Computing Center of NWPU are also gratefully acknowledged.

\end{document}